# Interfacial Phase Competition Induced Kondo-like Effect in Manganite-insulator Composites

Ling-Fang Lin[1], Ling-Zhi Wu[2], Shuai Dong[1][‡]

[1] Department of Physics, Southeast University, Nanjing 211189, China

[2] School of Geography and Biological Information, Nanjing University of Posts and Telecommunications, Nanjing 210046, China

Corresponding author. E-mail: [‡]sdong@seu.edu.cn

Kondo-like effect, namely the upturn of resistivity at low temperatures, has been observed in perovskite manganite when nonmagnetic insulators were doped as the secondary phase. In this paper, the low-temperature resistivity upturn effect was argued to be originated from the interfacial magnetic phase reconstruction. Heisenberg spin lattices have been simulated using Monte Carlo method to reveal the phase competition around the secondary phase boundary, namely the manganite-insulator boundary behaves a weak antiferromagnetic tendency. And the resistor-network model based on the double-exchange conductive mechanism indeed reproduces the low-temperature resistivity upturn effect. Our work provide a reasonable physical mechanism to understand novel transport behaviors in micro-structures of correlated electron systems.

## 1 Introduction

Perovskite manganites have attracted much attention owing to their strongly correlated electron characteristics, especially the colossal magnetoresistance (CMR) effect [1, 2]. The physical properties of manganite sensitively depend on spin and charge/orbital orders. And the competitions among various phases which have close free energies but divergent physical properties are responsible to the CMR effect [3, 4]. Despite the colossal effect, the CMR has not been successfully used in applications to replace the giant magnetoresistance (GMR) effect which is observed in magnetic metal multilayers. An important bottleneck of CMR is that it requires large magnetic fields, usually in the order of 1 Tesla [2], which is difficult to be realized in micro-devices.

To overcome this drawback, in recent years, researchers studied various manganite-insulator composites to improve the low-field magnetoresistance [5, 6]. With insulators embedded in wide-bandwidth manganite like $La_{2/3}Sr_{1/3}MnO_3$, prominent magnetoresistance can be realized with relative low fields. Not only the enhanced low-field magnetoresistance but also the Kondo-like effect, namely the upturn of resistivity at low temperatures ($T$'s), has also been observed when nonmagnetic insulators as a secondary phase were doped into La$_{2/3}$Sr$_{1/3}$MnO$_3$ [5, 7, 8]. Originally, the Kondo effect describes a characteristic electrical resistivity changing with temperature which results from the scattering of conduction electrons in a metal due to magnetic impurities [9].

However, the Kondo-like effect observed in manganite-insulator composites is somewhat different from the traditional Kondo effect for the following reasons. First, the carriers in manganite are spin-polarized and the non-magnetic insulators can't play as a magnetic scatter center as in Kondo effect. Second, the scale of non-magnetic cluster here is in the micro-meter scale, which is much larger than the mean free path of electrons in manganite, conceptually different from the atomic scale impurities in metal alloys. Extensive experimental and theoretical works [10-13] have been performed in past years, and several mechanisms have been proposed to interpret these resistivity minima, most of which are still under debate. A clear and wide-accepted physical scenario remains absent.

The double-exchange model proposed by Zener and a strong electron-phonon interaction arising from the Jahn-Teller splitting of Mn $3d$ levels can explain many of the electrical and magnetic properties of manganite. However, such a quantum model can't be directly applied to describe a transport issue in the micro-meter scale. Therefore, the mechanism of the resistivity minimum in manganite-insulator composites has not been completely understood [14-16].

In this article, we propose a new physical scenario based on the interfacial phase competition as the origin of the Kondo-like resistivity upturn effect. Due to the inherent phase competition between ferromagnetic phase and charge-ordered antiferromagnetic phase [1], previous studies have proved that the ferromagnetism of manganites can be seriously suppressed in the interfacial layers contacting insulators [17]. Meanwhile, the secondary phase boundary can lead to weak antiferromagnetic tendency [17]. Even though, the transport properties with such phase competed interfaces, have not been theoretically investigated.

## 2  Model and method

In the current work, the double-exchange conductance is calculated with classical resistor-networks for a Heisenberg spin model to reveal the fascinating Kondo-like effect in manganite-insulator composites. First, the Heisenberg spin model is a general model for magnetic systems and can be applied to describe the magnetic transition in La$_{2/3}$Sr$_{1/3}$MnO$_3$. Second, the resistor-network method has been successfully adopted to simulate the transport in manganite, which can handle a large-scale system without losing the double-exchange feature [18-23].

In our model, as show in Fig. 1, an isolated cluster of insulator (region III) is embedded in the magnetic manganite background (region I) as a secondary phase. Due to the interfacial reconstruction, such as the change of local electron concentration or strains, the manganite-insulator boundary (region II) will behave a weak antiferromagnetic (AFM) tendency with lower critical $T$'s than the original ferromagnetic (FM) Curie temperature ($T_C$) of the manganite itself [17].

The Heisenberg spin model on a two-dimensional (2D) square lattice ($L \times L$, $L = 31$ here) with periodic boundary conditions is adopted in the following simulation. The Hamiltonian of Heisenberg spin lattice model reads as:

$$H = -\sum_{\langle i,j \rangle} J_{i,j} S_i \cdot S_j - S_i \cdot h, \quad (1)$$

where $J_{i,j}$ is the exchange interaction between the nearest-neighbor (NN) spins $S_i$ and $S_j$, $h$ is the external magnetic field and the length of spin is normalized as unit 1. As shown in Fig. 1, $J_I$ is uniformly positive for the FM manganite, which is taken as the energy unit 1 for simplicity. For the interfacial region, $J_{II}$ is negative for weak AFM coupling, which's absolute value is smaller than $J_I$.

The standard Markov chain Monte Carlo (MC) method with the metropolis algorithm is employed to study the $T$-dependent properties of the lattice. In our MC simulation, the spin lattice is initialized randomly, then the first $1 \times 10^4$ MC steps (MCSs) are used for thermal equilibrium, and the following $1 \times 10^4$ MCSs are used for measurements. In all simulations, the acceptance ratio of MC updates is well controlled to be about 50% by adjusting the updating windows for spin vectors, which can give rise to the most efficient MC sampling [24].

In each MC measurement, a resistor-network is built based on the spin lattice and the total resistance ($R$) of the system is calculated [20-23]. As show in Fig.2, a resistor is assigned between each NN site pair with a value:

$$C_{i,j} = \sqrt{(1 + S_i \cdot S_j)/2}. \quad (2)$$

This formula can describe a spin-dependent double-exchange process. The physical meaning is that when adjacent spins are parallel, the conductance is maximal, and conversely when NN spins are antiparallel, the conductance is forbidden. Between these two limits, the conductance is determined by the spin angle according to the double-exchange formula [25]. The conductance between the regions I and II is set to be 0, as well as within in the region I. The whole resistor network is solved by using Kirchhoff's equations. In detail, a system of linear Kirchoff's equations is constructed for each site, which reads:

$$\sum_j C_{i,j}(V_i - V_j) = 0, \quad (3)$$

where $V_i$ and $V_j$ are the voltages at NN sites $i$ and $j$, respectively; $C_{i,j}$ is the conductance of local resistor between sites $i$ and $j$, calculated by Eq. 2.

It should be noted that although the resistor-network has been used to study the percolation in phase separated manganite, the present model goes beyond previous studies [20-23].

Previously, two or three types of fixed resistors are used to construct the network, which can't simulate the interfacial effect in the region II. Only when considering the detail correlation between Heisenberg spins, as done in the current work, the model can handle the spin-dependent transport in such manganite-insulator composites.

## 3 Results and discussion

First, the $T$-dependent of $R$'s are compared between the pure FM bulk (with region III but no region II) and the one with the interfacial region II, as shown in Fig. 2(a). For the pure FM one, the $R$-$T$ curve displays a good metallic behavior, namely $R$ decreases in the cooling progress, even the lattice has an insulator embedded in. When the interfacial region II is taken into account, the upturn of resistivity indeed show up obviously at low $T$'s and the minimal $R$ appears at $T=0.6$ when the doping concentration is 20.5% characterized by the volume of region III ($V_{III}$). In above simulation, the weak exchange interaction in region II is $J_{II}$=-0.4 and the thickness of region II is $d$=4. To further understanding the underlying mechanism, one more simulation is done by setting $J_{II}$ as 0 and keeping all other parameters unchanged. In this case, the spins in region II are always in random directions (paramagnetic), and the upturn resistivity effect disappears, as shown in Fig. 2(a). Above comparisons imply the key role of AFM coupling in interfacial region II in the Kondo-like resistivity upturn.

To understanding the relationship between magnetism and transport behavior, two MC snapshots of the spin lattice at $T=0.05$ and $T=0.4$ are shown in Fig. 2(b) and 2(c) respectively. In both cases, the lattice displays a robust FM order in region I. In region II, spins are somewhat "disordered" at $T=0.4$, and the "disordered" pattern gradually turn to be AFM ordered at low $T$'s.

The direct relationship between this magnetic transition and upturn of $R$ can be intuitively visualized by the profiles electrical potential differences. Fig. 2(d) shows the contour map of potential difference of the whole resistor network between the pure FM lattice and the one with region II at $T=0.05$. Similarly, Fig. 2(e) shows the potential difference between $T=0.05$ and $T=0.4$ for the same lattice with region II. In both contour maps, there is a distinct difference surrounding the embedded insulator cluster, namely the interface layers between manganite and insulator.

As a summary of Fig. 2, the manganite matrix, which changes from paramagnetic to FM order with deceasing $T$, gives a metallic transport behavior. Then the AFM phase transition in region II is responsible for the resistivity upturn in the low $T$'s considering the double-exchange process as presented in Eq. (2).

Above simulation has shown a possible physical process to induce the Kondo-like resistivity upturn effect. In the following, the magnetic field and temperature-dependent of the resistivity has been systematically simulated. First, the different doping concentrations are simulated by changing the relative volume $V_{III}$ for a fixed thickness of region II ($d$=5) and weak exchange interaction in region II ($J_{II}$=-0.4). As shown in Fig. 3(a), the upturn amplitude of $R$'s is more prominent with increasing $V_{III}$ while the temperatures of $R$'s minimums ($T_{min}$) are almost unshift. It is reasonable since the interfacial perimeter (and thus the volume of region II) increases linearly with the radius of embedded insulator. Second, by fixing the values of d and $V_{III}$, T of minimal resistance shows an increasing tendency with the absolute value of $J_{II}$, as displayed in Fig. 3(b). The reason is also simple: a stronger $J_{II}$ corresponds with a higher AFM transition $T$. Even though, the amplitude of upturn seems to be stronger in the weak $|J_{II}|$ case. Third, Fig.3(c) shows the changes of $R$-$T$ curves with increasing thickness of region II while other parameters are fixed. The value of $T_{min}$ is almost unchanged while the upturn amplitude is greatly enhanced in the thick $d$ case. Last, we switch on the external magnetic field and study the magnetoresistance effect. As shown in Fig. 3(d), it is obvious that the applied magnetic field suppresses the upturn effect. The value of $T_{min}$ gradually decreases with the increasing field and finally the system turns to be a full metal. This process can be understood as the suppression of weak AFM correlation in region II by magnetic fields, which polarizes all spins parallel and thus the whole FM correlation is established in region I and II. <u>In our simplified model, the $T_{min}$ corresponds to the interfacial magnetic phase transition, which is determined by the strength of the exchange interactions $J_2$ and external magnetic field $h$. Therefore, $T_{min}$ is almost unchanged upon $d$'s increasing, but can be shifted by applied magnetic field.</u>

Our above results show that the low-$T$ resistivity upturn can be tuned by the interfacial parameters as well as external magnetic fields. The amplitude of upturn and $T_{min}$ depends on different physical factors. Our study can help researchers to fine tune the Kondo-like effect in future experiments.

## 4  Summary

In conclusion, Monte Carlo simulations have been performed on Heisenberg spin lattices and resistor-network model to qualitatively investigate the low-temperature resistivity upturn effect in manganite-insulator composites. The resistor-network is constructed based on the spin-dependent double-exchange process. Temperature-dependent of the resistivity at different conditions has been systematically simulated. Our simulation found that the antiferromagnetic tendency of manganite-insulator boundary due to the interfacial reconstruction can lead to the low-temperature resistivity upturn effect. Our toy model provided a simple but intuitive physical picture to clarify a possible mechanism of Kondo-like transport behavior in correlated electronic materials.

**Acknowledgements** We thank G. X. Cao, Y. Z. Gao, Y. K. Tang for helpful discussions. This work was supported by the Natural Science Foundation of China (Nos. 11274060 & 51322206).

**Figures**

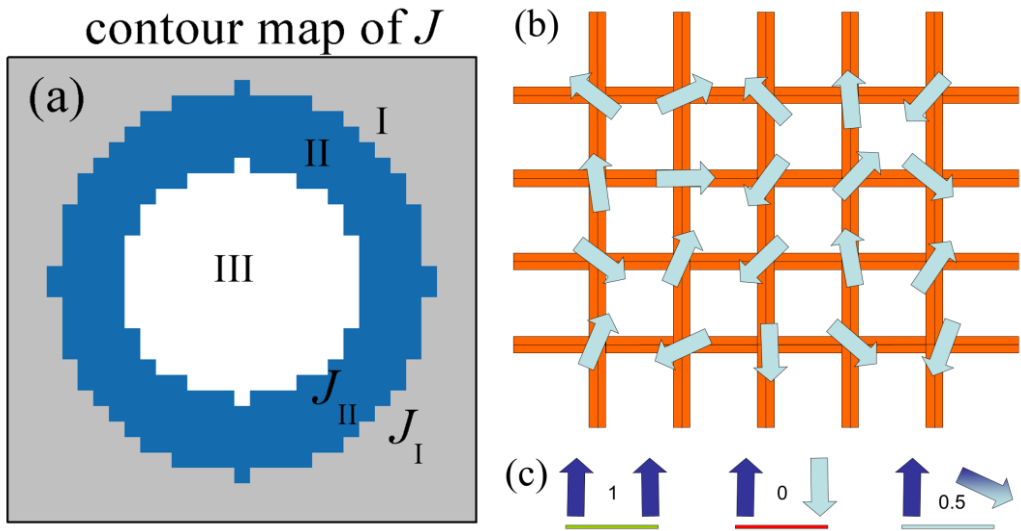

**Fig. 1 (a)** Schematic diagram of the contour map of *J* on the 2D spin lattices. Region I, II, and III represent the manganite background, manganite-insulator boundary and insulating impurity cluster, respectively. **(b)** Schematic diagram of the network resistor of the Heisenberg spin lattice with periodic boundary conditions. The arrow on each site represents spin. The orange frame connecting every NN sites is the resistor-network. **(c)** The sketch of local resistor calculated based on the double-exchange process between NN spins pair. Left: when adjacent spins are parallel, the local conductance is maximal 1. Middle: when NN spins are antiparallel, the local conductance is 0. Right: for arbitrary angle between NN spins pair, for example $120^\circ$, the local conductance can be calculated by using Eq. 2.

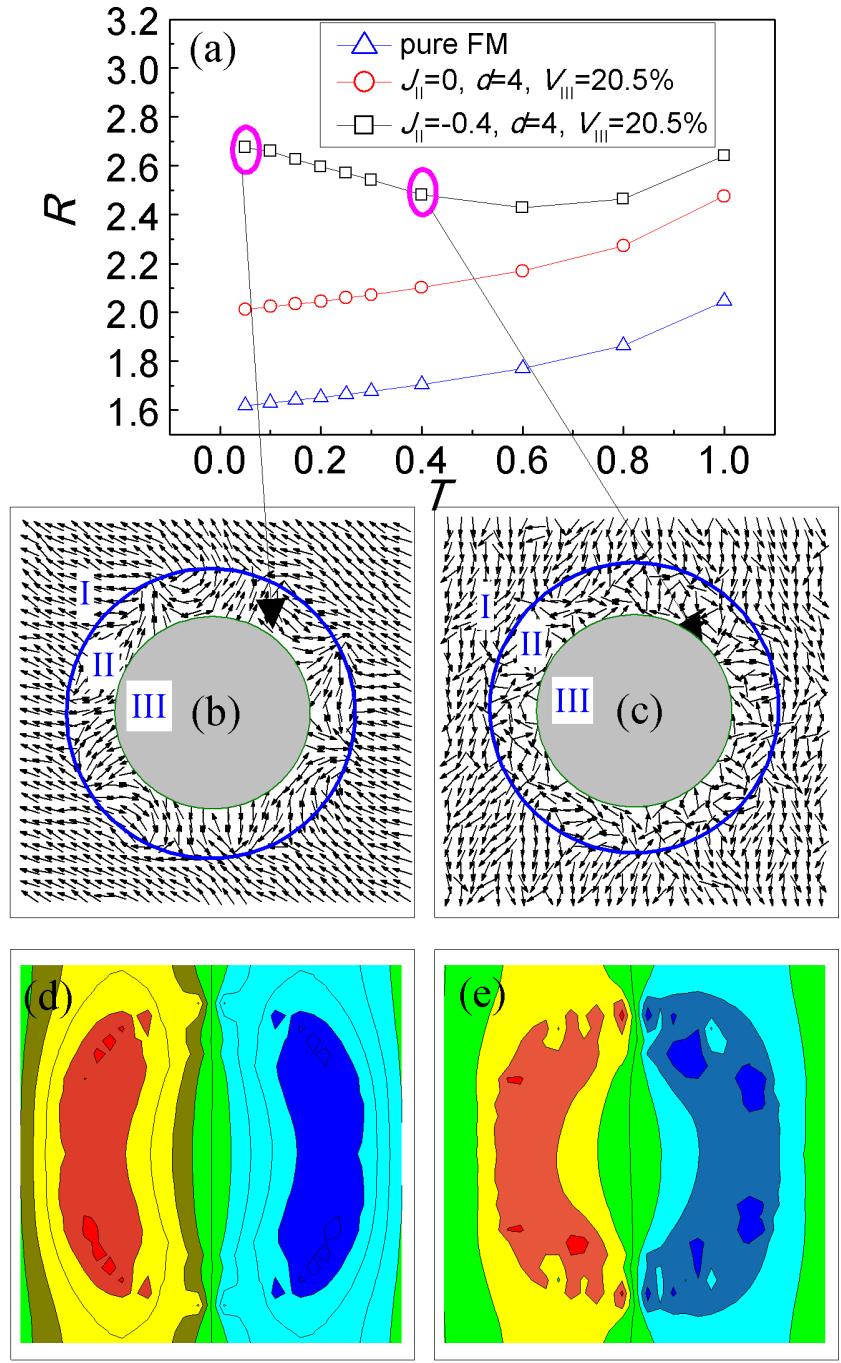

**Fig. 2 (a)** MC results of $T$-dependent of $R$. It is a comparison between these cases: 1) $J_{II}=J_{I}=1$, a pure FM lattice; 2) $J_{II}=0$, the one with disordered interfacial layers; 3) $J_{II}=-0.4$, the regular one with interfacial AFM coupling; **(b)**-**(c)** Snapshots of the 2D spin patterns at $T=0.05$ and $T=0.4$, respectively, corresponding to the circles in Fig.2(a). **(d)**-**(e)** Contour maps of potential difference of the whole resistor-network: (d) between 1) the pure FM lattice and 2) the regular one at $T=0.05$; (e) between 1) $T=0.05$ and 2) $T=0.4$ for the identical regular lattice.

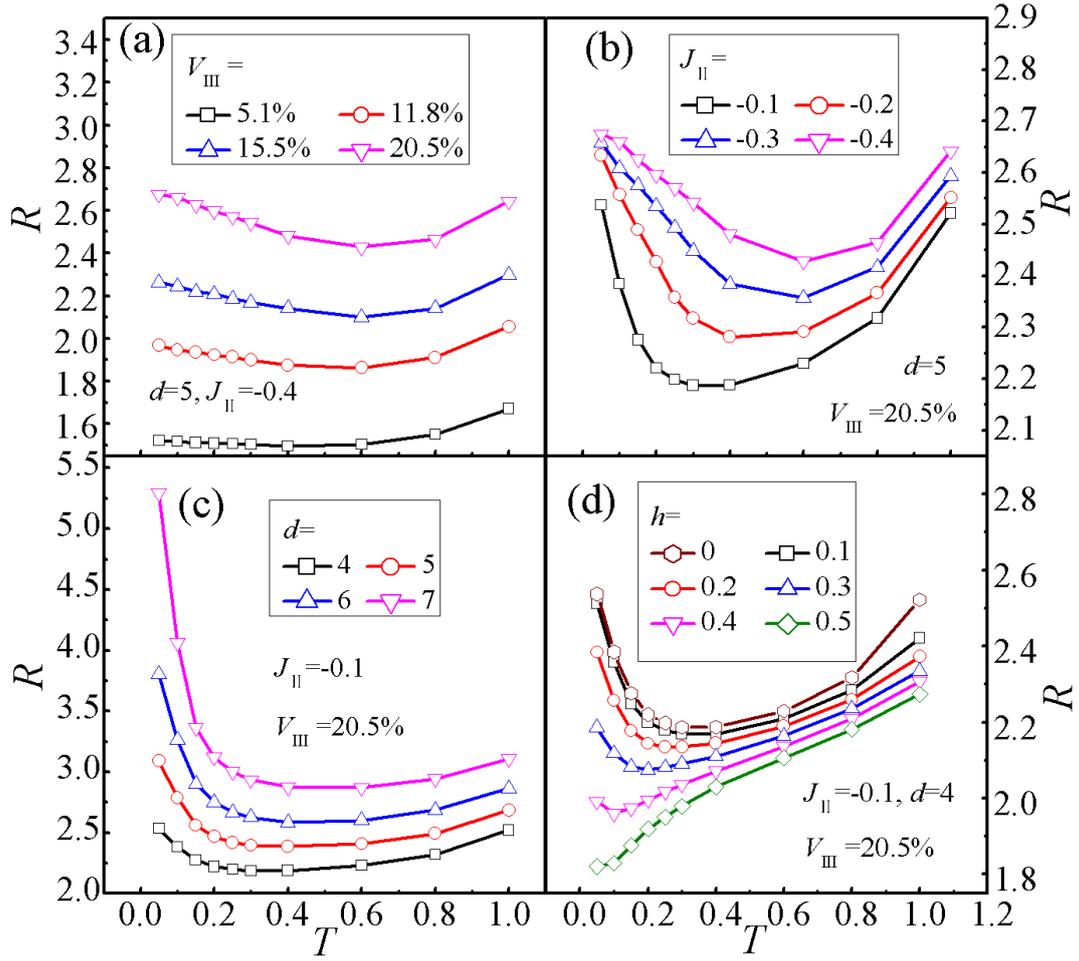

**Fig. 3** MC results of resistance as a function of $T$ at **(a)** different doping concentrations characterized by the volume of insulator cluster; **(b)** exchange interactions in region II ($J_{II}$); **(c)** thickness of region II; **(d)** external magnetic field $h$.